\documentstyle[preprint,floats,aps,epsf,prb]{revtex}

\newcommand \be  {\begin{equation}}
\newcommand \ee {\end{equation}}
\newcommand \bea {\begin{eqnarray} }
\newcommand \eea {\end{eqnarray}}

\begin{document}
\draft
\title{Quantum Lifshitz Point}
\author{R. Ramazashvili}
\address{
Loomis Laboratory, University of Illinois at Urbana-Champaign, \\ 
Urbana, IL 61801-3080, USA. }
\maketitle
\date{\today}
\maketitle
\begin{abstract}
I study a quantum Lifshitz point in a three-dimensional 
itinerant antiferromagnet, in particular the scaling  
of the N\'{e}el temperature, the correlation length, 
the staggered susceptibility, the specific heat 
coefficient and the resistivity. At low temperatures, 
the model is shown to have the inverse staggered 
susceptibility and the resistivity varying as T$^{5/4}$, 
and the specific heat coefficient varying as T$^{1/4}$. 
\end{abstract}
\vskip 0.2 truein
\pacs{PACS numbers:}


\section{Introduction}

Magnetic metals close to a zero temperature phase transition
have been a subject of active ongoing study \cite{santa-barbara}, 
mainly due to their apparent defiance of the Fermi liquid paradigm. 
Near a quantum critical point, one usually finds neither a nearly 
constant linear specific heat coefficient 
$\gamma \equiv \delta C / T \sim const.$, 
nor a virtually temperature-independent magnetic susceptibility 
$\chi \sim const.$, 
nor a $T^2$-resistivity $\rho \sim const. + T^2$,
all normally expected from a Landau Fermi liquid \cite{landau1}. 
Instead experiments reveal a host of anomalous thermodynamic, 
magnetic and transport properties \cite{santa-barbara} calling 
for adequate theoretical description. 

In its current form, scaling theory of quantum criticality 
in itinerant magnets is due to Hertz \cite{hertz1} 
and Millis \cite{millis1}.  
It describes incipient magnetic ordering in 
an isotropic metal solely in terms of a boson order parameter 
with the correlation range in both space and time 
diverging at a quantum critical point. 
This divergence naturally leads to low-temperature anomalies 
in thermodynamic, magnetic and transport properties. 
Some of the early work on the subject was done by Mathon 
\cite{mathon1}, Makoshi and Moriya \cite{makoshi1}, Ueda \cite{ueda1} 
and Dzyaloshinskii and Kondratenko \cite{denko1}, 
who described magnetic transitions 
at $T=0$ using diagrammatic methods. These and other 
studies of nearly ferro- and antiferromagnetic metals as well 
as magnetic fluctuations in general, done in the spirit 
of self-consistent theory, were summarized 
in the book by Moriya \cite{moriya1}. A phenomenological 
approach aiming at establishing relations between various 
critical exponents at a quantum critical point has been 
developed by Continentino \cite{continentino1}. 

The Hertz-Millis theory and its implications describe 
rather well some itinerant ferromagnets such as ZrZn$_2$ 
and, to a lesser extent, MnSi. In ZrZn$_2$, moderate external 
pressure $p \approx$ 8 kbar is enough to reduce the Curie temperature 
$T_C$ to zero. Near the critical pressure $p_c$, 
$T_C$ scales as $(p_c - p)^{3/4}$ while the resistivity 
obeys \cite{ogawa1,julian1} $\rho \sim T^{1.6\pm0.1}$, 
in excellent agreement with 
the theory \cite{millis1,mathon1,moriya1}.

At the same time, several Ce-based antiferromagnets, 
where the N\'{e}el temperature can be tuned to zero 
by pressure or doping, appear to disregard the conclusions
of the available theories. One of the most studied examples
is \cite{lohneysen1} CeCu$_{6-x}$Au$_x$, 
where at the quantum critical point 
the resistivity behaves as $const. + T$, and where 
over more than two decades in temperature the specific heat 
coefficient diverges as $\ln(T)$ while the inverse 
susceptibility fits \cite{schroeder2} 
$\chi^{-1} \sim const. + T^{\alpha}$
with $\alpha \approx 0.8 \pm 0.1$. 

Another interesting case is CePd$_2$Si$_2$, a metal which orders 
antiferromagnetically below 10 K. Applied pressure 
reduces the N\'{e}el temperature $T_N$ from 10 K at the 
ambient pressure 
to about 0.4 K at 28 kbar. 
Between 15 kbar and 28 kbar, $T_N$ falls linearly with pressure, 
and at 28 kbar CePd$_2$Si$_2$ 
becomes superconducting below 0.4 K, in a narrow strip
between 23 and 32 kbar 
\cite{julian1}. 
At 28 kbar, the resistivity of CePd$_2$Si$_2$ 
exhibits a striking $T^{1.2\pm0.1}$ behavior over about 
two decades in temperature 
\cite{julian1}. 

Neither CeCu$_{6-x}$Au$_x$ nor CePd$_2$Si$_2$ appear to respect 
the available theoretical results which, for a bulk 
antiferromagnet, would imply the N\'{e}el temperature scaling 
\cite{millis1} as 
$T_N \sim (p_c - p)^{2/3}$, the specific heat coefficient 
\cite{makoshi1,moriya1} 
$\gamma \sim T^{1/2}$ and the resistivity \cite{ueda1,moriya1} 
$\rho \sim T^{3/2}$. 
At the same time, theoretical results for {\it two} dimensions 
are much closer to what was observed in CeCu$_{6-x}$Au$_x$, 
as in this case the theory does yield $const. + T$ resistivity \cite{rosch1}, 
logarithmically divergent linear specific heat coefficient
and essentially linear scaling of the N\'{e}el temperature 
with pressure \cite{millis1}. This led to an interpretation of data 
on CeCu$_{6-x}$Au$_x$ in terms of a purely two-dimensional 
magnetic ordering in an otherwise perfectly three-dimensional 
metal with only moderate anisotropy \cite{rosch1,stockert2}. 

An alternative interpretation 
\cite{schroeder2}, based on the neutron scattering data,  
pointed at a possible explanation in terms 
of highly anisotropic 
critical fluctuations with the stiffness in one of the directions 
vanishing at or very close to the quantum critical point. 
The conspiracy of this anisotropy with the anomalous frequency 
exponent \cite{schroeder2} $\alpha \approx 0.8 \pm 0.1$ 
of critical fluctuations allowed to fit together the large body 
of experimental data 
obtained at the quantum critical point in CeCu$_{6-x}$Au$_x$, 
including the specific heat, the uniform susceptibility and the 
neutron scattering scans.

This latter scenario would imply a residual quartic dispersion 
($\sim q^4$) of critical fluctuations 
in the ``soft'' direction, thus also leading 
to the dimensionality reduction (by $1/2$, as opposed 
to the purely two-dimensional order) 
and hence to qualitative modification of the Hertz-Millis theory. 

Finally, it was  noticed \cite{grosche1,mathur1} that the resistivity 
$\rho \sim T^{1.2\pm0.1}$ of CePd$_2$Si$_2$, as well as 
$\rho \sim T^{1.25\pm0.1}$ of CeNi$_2$Ge$_2$ (which has the 
structure of CePd$_2$Si$_2$ with a smaller unit cell) may also be 
explained assuming anisotropic critical fluctuations with the 
residual quartic dispersion in one of the directions.  

All these experimental findings suggest that 
vanishing stiffness may be the aspect of physics 
needed to to describe the quantum criticality in Ce-based 
antiferromagnets by a Hertz-Millis type of theory. 
A critical point which, in addition to the onset of ordering, 
is characterized by disappearance of stiffness 
in one or several directions, is called a Lifshitz point \cite{lifshitz1}. 
In this paper, I study a {\it quantum} Lifshitz point, 
a curious yet possibly experimentally relevant coincidence 
of a {\it quantum} critical point (onset of ordering at $T = 0$) 
with a point where the stiffness vanishes 
in one or several directions in the momentum space.
I develop scaling theory 
of a classical Gaussian region \cite{millis1} 
of a disordered phase near a quantum Lifshitz point 
in an itinerant three-dimensional magnet. 
I study a particular case of a Lifshitz point, 
where the incipient ordering exhibits anisotropic dispersion 
\cite{schroeder2,grosche1,mathur1} which is quartic in only 
{\it one} direction and quadratic in the remaining two.

The restriction to only the classical Gaussian region 
is due to the fact that in the other regions the low-temperature 
behavior is dominated by crossovers between various regimes 
and thus comparison to the experiment is rather hard to make. Moreover, 
the full phase diagram is sensitive to the relative strength of the 
coupling constants, whereas the results for the classical Gaussian 
region of a disordered phase are independent of this uncertainty 
and can be tested experimentally. 

It is important to note that Lifshitz point is 
a multicritical point and that, generally, one shall expect 
to find three phases in its vicinity, corresponding to one disordered 
and two different ordered states (see Section V). 
At the same time, none of the experiments on CeCu$_{6-x}$Au$_x$ or
CePd$_2$Si$_2$ which I am aware of, indicated presence of more 
than one magnetically ordered phase. 
To that end, I show that, near a {\it quantum} Lifshitz point, 
one may indeed find only {\it one} ordered phase rather than two, 
which makes the model potentially relevant to CeCu$_{6-x}$Au$_x$ 
\cite{lohneysen1,schroeder2,rosch1,stockert2}, 
CePd$_2$Si$_2$ and CeNi$_{2}$Ge$_{2}$ 
\cite{julian1,grosche1,mathur1,steglich1,fukuhara1,aoki1}. 

Bearing in mind that both 
CePd$_2$Si$_2$ and CeCu$_{6-x}$Au$_x$ are
antiferromagnetic metals, one is led to assume 
that the critical mode 
is non-conserved and over-damped. Therefore this work may be 
viewed as an experiment-motivated extension of the Hertz-Millis theory 
\cite{hertz1,millis1} to a particular case of a quantum Lifshitz point. 
 
In Section II, I present the results and discuss them.
I determine the equation of the critical line, 
the behavior of the order parameter susceptibility 
and the correlation length close to the quantum Lifshitz point.
I also estimate the low-temperature behavior of the conductivity
and the anomaly of the linear specific heat coefficient.
Then I compare these theoretical results with 
the experimental findings 
\cite{julian1,stockert2,grosche1,steglich1,fukuhara1,aoki1}. 

In Section III, I derive the scaling equations,
solve them in Section IV and obtain the results 
outlined in Section II.
In Section V, I discuss the phase diagram
near the quantum Lifshitz point and set 
the conditions for appearance
of only one ordered phase. 

To demonstrate that the quantum Lifshitz point I study 
is well defined, in Section VI I examine how interactions 
generate stiffness in the ``soft'' direction, and show 
that this effect can be neglected at all relevant momenta 
and frequencies. Finally, in Section VII, I summarize 
the results and comment on them. 
The appendix provides the calculation details. 

Self-consistent treatment of a model with the stiffness 
vanishing in one of the directions in the momentum 
space has been given recently by C. Lacroix {\it et al.} 
\cite{lacroix1} 
with reference to the experimental data \cite{ballou1} 
on YMn$_2$-based materials, where the transition is {\em first-order}. 
The present work amounts to a renormalization group 
derivation of some of the results obtained in \cite{lacroix1} 
(staggered susceptibility), plus new results 
(specific heat coefficient, possible phase diagrams 
and the transition line equation) 
along with comparison to the experimental 
data for Ce-based metallic antiferromagnets undergoing 
second-order phase transition. It is also shown that 
generation of stiffness in the ``soft'' direction 
by short-distance fluctuations is negligible at all relevant 
momenta, and thus the quantum Lifshitz point studied in this 
paper is well-defined.

\section{The results}

The main results of this Section amount to establishing 
the leading low-temperature behavior of various quantities 
near a quantum Lifshitz point. This task is facilitated by 
the fact that the theory falls above its upper critical 
dimension. Moreover, since the stiffness in the ``soft'' direction, 
generated by the short-range fluctuations, turns out 
to be negligible (see Section VI), the thermodynamic and 
transport properties in the classical Gaussian region 
($T_N = 0$) may be obtained from the Gaussian action 
with this stiffness set equal to zero: 
\be  
S[ \phi ] = \int_{0}^{\beta} d\tau \int dx 
\phi_{\alpha}
\left[
\delta + 
|\partial_\tau| + \nabla^{2}_{\parallel}
 + \nabla^{4}_{\perp}
\right]
\phi_{\alpha},  
\label{eq:simple}      
\ee
where $\delta$ is defined by the tuning parameter $p$ and by
the feedback of the quartic interaction 
in the Ginzburg-Landau action (\ref{eq:action1}-\ref{eq:action3}) 
as per $\delta= (p - p_c)/p_c + const. T^{5/4}$. 
Thus the N\'{e}el temperature $T_N$ in this theory scales as  
$T_N \sim (p_c - p)^{4/5}$. At $p = p_c$, the correlation length 
$\xi_{\parallel}$ scales as $T^{-5/8}$, whereas the correlation length 
$\xi_{\perp}$ in the ``soft'' direction scales as $T^{-5/16}$.

The leading exponent 
of the specific heat coefficient is given by the
Gaussian contribution to the free energy $F$:
$$F = Tr \log G^{-1} (q, \omega) = 
\int^1 dz \int^1 d^{2}q_{\parallel} \int^1 dq_{\perp} 
\coth{\frac{z}{2T}} \arctan \left[ \frac{z}{q_{\parallel}^2 
+ q_{\perp}^4} \right] $$  
and yields the specific heat coefficient 
$$\delta C/T \sim T^{1/4}.$$

Resistivity due to scattering off 
anisotropic fluctuations can be estimated via the characteristic 
transport time $\tau_{tr}$ given by the fluctuations of the 
magnetic order parameter near the transition. Since the critical 
fluctuations are antiferromagnetic and thus characterized by 
a finite wave-vector, the $(1 - \cos \theta)$ factor in the transport 
relaxation rate may be omitted in a qualitative estimate, 
which leads to  $$1/\tau_{tr} \sim T\sum_{q, \omega} 
\langle \phi^2 \rangle_{q, \omega} \sim T^{5/4},$$ 
as realized indeed by the authors of \cite{grosche1,mathur1}. 
Recently, it has been argued \cite{rosch7} that the observed resistivity 
$\rho(T) \sim const. + T^{1.2\pm 0.1}$ may be a crossover 
phenomenon due to the interference between the impurity scattering 
and scattering by the critical fluctuations, conspiring to mimic 
a power-law behavior. At the moment it remains to be seen how sensitive 
this mechanism may be to the material-dependent factors such as the 
shape of the Fermi surface. At the same time, for a system with a true 
quantum Lifshitz point, the $T^{5/4}$ resistivity shall be accompanied 
by the $C/T \sim T^{1/4}$ scaling of the specific heat coefficient. 


In CePd$_{2}$Si$_{2}$ and CeNi$_{2}$Ge$_{2}$, the resistivity does exhibit 
\cite{julian1,grosche1} a temperature exponent close to $5/4$. 
However, the rest of the available data is less encouraging for the 
present theory. 
The N\'{e}el temperature scales linearly with pressure 
\cite{julian1} instead of obeying 
$T_N \sim (p_c - p)^{4/5}$. To my knowledge, 
the specific heat data on CePd$_{2}$Si$_{2}$ at the critical 
pressure is not yet available. The specific heat data 
on CeNi$_{2}$Ge$_{2}$ is still ambiguous, as both the $C/T \sim T^{1/2}$ 
and $C/T \sim \log T$ have been reported \cite{steglich1,aoki1}. 

One can indeed find formal reasons (such as presence of 
the superconducting phase in CePd$_{2}$Si$_{2}$, or disorder 
in CeCu$_{5.9}$Au$_{0.1}$) 
why the Lifshitz point theory is not applicable to these materials. 
However, the $T^{1.2}$ scaling of the resistivity in CePd$_{2}$Si$_{2}$ 
is seen at temperatures up to 60 K, which is more than 
two orders of magnitude 
greater than the superconducting transition temperature 
$T_c \approx$ 0.4 K. At the same time, virtually linear scaling 
of $T_N$ with pressure persists up to about 5 K, 
which is again much greater than $T_c$. 
These arguments (as well as recent finding 
\cite{schroeder2}
of the $\omega/T$ scaling and the inverse {\it uniform} 
susceptibility behaving as $\chi^{-1} \sim const. + T^{0.8 \pm 0.1}$ 
in CeCu$_{5.9}$Au$_{0.1}$) strongly suggest that 
the physics of these materials amounts to more 
than just a refined Gaussian theory 
above its upper critical dimension.  

\section{The scaling equations}

The assumption of an overdamped critical mode
with the stiffness vanishing in one of the directions 
leads to the following effective action close to the quantum 
Lifshitz point:
\bea
S_{eff}[ \phi ] &=& S^{(2)} + S^{(4)}         \label{eq:action1}\\
S^{(2)}         &=& 
\int_{0}^{\beta} d\tau \int dx 
\phi_{\alpha}
\left[
\delta + 
|\partial_\tau| + \nabla^{2}_{\parallel}
 + D \nabla^{2}_{\perp} + 
\nabla^{4}_{\perp}
\right]
\phi_{\alpha}                                 \label{eq:action2}\\                                        
S^{(4)}        &=& 
\int_{0}^{\beta} d\tau \int dx 
\left[
u (\phi_{\alpha}\phi_{\alpha})^2
+ v_1 (\nabla_{\perp}\phi_{\alpha})^2 \phi_\beta \phi_\beta + v_2 
(\phi_{\alpha} \nabla_{\perp} \phi_{\alpha})^2
\right].                                     \label{eq:action3} 
\eea
with the frequency and the momentum cut-off set equal to unity. 
Here the mass term $\delta$ and the stiffness $D$ 
in the ``soft'' direction describe deviation 
from the quantum Lifshitz point, 
$\beta=1/T$ is the inverse temperature, 
$\nabla_{\parallel}$ corresponds to the two ``normal''
directions, whereas $\nabla_{\perp}$ corresponds to the 
third, ``soft'' direction. The coupling constants 
$v_1$ and $v_2$ describe the dispersion of the quartic 
coupling constant $u$; they correspond to the only two 
possible 
linearly independent terms quadratic in $\nabla_\perp$ 
and quartic in $\phi_\alpha$, and describe generation 
of stiffness in the ``soft'' direction, 
as shown in Fig. \ref{fig:stiffness}. 

The scaling variables in the theory are the mass term $\delta$,
the stiffness $D$ in the ``soft'' direction, the 
temperature $T$ and coupling constants $u$, $v_1$ and $v_2$. 
To the lowest order in the latter three, the scaling equations
can be derived e.g. by expanding the partition function
$Z$ to first order in $S^{(4)}$, then 
integrating out a thin shell near the cut-off, 
rescaling the variables and the fields, and then comparing the 
result with the original action \cite{millis1}: 
\bea
\frac{d T(b)}{d \ln b}      &=&  2  T(b)             \label{eq:t}\\
\frac{d u(b)}{d \ln b}      &=&  -  \frac{1}{2} u(b) \label{eq:u}\\
\frac{d v_{1, 2}(b)}{d \ln b} &=&  -  \frac{3}{2} v_{1, 2}(b) 
                                                     \label{eq:v}\\      
\frac{d \delta(b)}{d \ln b} &=&  
2  \delta(b) + 2 (n + 2) u f_{1}[T(b), \delta(b), D(b)] \nonumber\\
&+& (n v_1 + v_2)f_{2}[T(b), \delta(b), D(b)]         \label{eq:delta}\\
\frac{d D(b)}{d \ln b} &=& D(b) + (n v_1 + v_2) 
f_{1}[T(b), \delta(b), D(b)].                        \label{eq:d} 
\eea
Here $n$ is the number of components of field $\phi_{\alpha}$.
The definitions of $f_{1}[T(b), \delta(b), D(b)]$ 
and $f_{2}[T(b), \delta(b), D(b)]$, 
as well as the details of the derivation, are given in the Appendix. 
Note that, following Millis \cite{millis1}, I denote the running value 
of a scaling variable (e.g. $T(b)$) by indicating explicit dependence 
on the rescaling parameter $b$, whereas for the bare (physical) 
quantities the $b$ dependence is omitted. Also note that truncating 
the equations at the first order in $u$, $v_1$ and $v_2$ is equivalent
to the assumption that the bare value of these couplings is small.

An extra scaling variable $D(b)$ in equations (\ref{eq:t}-\ref{eq:d}) 
compared with those of Millis \cite{millis1} leads to appearance of 
{\it two} different ordered phases to be described in Section V. 
This sets the main distinction between a quantum Lifshitz point 
and a simple quantum critical point of Hertz and Millis, 
where there is only one ordered phase. 

\section{Solution of the scaling equations}

In this Section, I obtain the qualitative solution 
of the scaling equations (\ref{eq:t}-\ref{eq:d}) 
following Millis \cite{millis1} and noticing that 
$f_1[T(b), \delta(b), D(b)]$ and $f_2[T(b), \delta(b), D(b)]$ 
virtually do not depend on $\delta(b)$ or $D(b)$ 
for $\delta(b), D(b) \ll 1$,
and that both of them fall off rapidly as $\delta(b)$ 
exceeds unity. Thus one can neglect their dependence 
on $\delta(b)$ and $D(b)$ for $\delta(b) < 1$ and stop the scaling 
at $\delta(b) = 1$. 

With these provisos, the formal solution of (\ref{eq:t}-\ref{eq:d}) reads 
\bea
T(b)       &=&  T b^2                            \label{eq:solution-t}\\
u(b)       &=&  u b^{-1/2}                       \label{eq:solution-u}\\
v_{1, 2}(b) &=& v_{1, 2} b^{-3/2}                \label{eq:solution-v}\\
\delta(b)  &=&  
\delta b^2 + 2 b^2 (n + 2) u \int_{0}^{\ln b}
d \tau e^{-\frac{5}{2}\tau} f_{1}[T e^{2\tau}] \nonumber\\
 &+& b^2 (n v_1 + v_2) \int_{0}^{\ln b} d \tau 
e^{-\frac{7}{2}\tau} f_{2}[T e^{2\tau}] 
\label{eq:solution-delta}\\
D(b) &=& b \left[ D + (n v_1 + v_2) \int_{0}^{\ln b}
d \tau e^{-\frac{5}{2}\tau} f_{1}[T e^{2\tau}] \right]. \label{eq:solution-d}
\eea

As in the case of a simple quantum critical point \cite{millis1}, 
two major regimes exist, depending on whether the scaling 
stops when the running value of the temperature is much 
smaller or much greater than one. The first case is usually referred to
as quantum while the second case is called classical. 

To obtain the condition for the quantum regime, set $T=0$ 
on the right hand side of (\ref{eq:solution-delta}), then integrate
over $\tau$ up to $\ln(b^{\ast})$ such that $\delta(b^{\ast}) = 1$, 
solve for $b^{\ast}$, substitute $b^{\ast}$ into (\ref{eq:solution-t}) 
and require $T(b^{\ast}) \ll 1$. The resulting condition is 
\be
T \ll r_{1},  \qquad 
r_{1} \equiv \frac{4}{5}(n+2)uf_{1}[0] + \frac{2}{7}(nv_1+v_2)f_{2}[0].
                                  \label{eq:quantum-classical-crossover}
\ee
If reversed, the inequality (\ref{eq:quantum-classical-crossover}) 
corresponds to the classical regime, where it is convenient to divide
the scaling trajectory into two parts, corresponding to $T(b) \ll 1$ 
(quantum) and $T(b) \gg 1$ (classical). 
For $T(b) \gg 1$, one can estimate $f_{1, 2}[T]$ 
as  $f_{1}[T] \approx B_1 T$, $f_{2}[T] \approx B_2 T$. Then, for 
$T(b) \gg 1$, the equations (\ref{eq:u}-\ref{eq:d}) 
may be recast in terms of new variables 
$U(b) \equiv u(b) T(b)$, $V_{1,2}(b) \equiv v_{1,2}(b) T(b)$: 
\bea
\frac{d T(b)}{d \ln b}      &=&   2  T(b)             \label{eq:t1}\\
\frac{d U(b)}{d \ln b}      &=&   \frac{3}{2} U(b) \label{eq:u1}\\
\frac{d V_{1, 2}(b)}{d \ln b} &=& \frac{1}{2} V_{1, 2}(b) 
                                                     \label{eq:v1}\\      
\frac{d \delta(b)}{d \ln b} &=&  
2  \delta(b) + 2 B_1 (n + 2) U(b) 
 + B_2 (n V_1(b) + V_2(b))                        \label{eq:delta1}\\
\frac{d D(b)}{d \ln b} &=& D(b) + B_1 (n V_1(b) + V_2(b)). 
                                                       \label{eq:d1} 
\eea

The initial conditions correspond to $b=\bar{b}$ 
such that $T(\bar{b})=1$ and read
\bea
T(\bar{b}) &=& 1         \label{eq:solution-t-bar} \\
U(\bar{b}) &=& u(\bar{b}) = u T^{1/4} \label{eq:solution-u-bar}  \\
V_{1,2}(\bar{b}) &=& v_{1,2} T^{3/4}  \label{eq:solution-v-bar}  \\
\delta(\bar{b}) &=& \frac{1}{T} \left[ r_1 + A (n+2) u T^{5/4}\right]
                                      \label{eq:solution-delta-bar} \\
D(\bar{b}) &=& \frac{1}{\sqrt{T}} 
\left[ r_2 + \frac{A}{2} (nv_1+v_2) T^{5/4} \right], 
                                      \label{eq:solution-d-bar}
\eea
where $r_2$ is defined by 
\be
r_2 \equiv D + \frac{2}{5}(nv_1+v_2)f_1[0].
\ee
For $T \gg r_{1}$, both 
$U(\bar{b})$ and $V_{1,2}(\bar{b})$ are small, which justifies 
using linearized equations near $b=\bar{b}$. Neglecting the higher 
powers of $T$ and assuming that $u$ and $v_{1,2}$ are of the 
same order of magnitude, the solution of (\ref{eq:t1}-\ref{eq:d1}) is
\bea
T(b)  &=& T(\bar{b}) b^2                  \label{eq:solution-t1} \\
U(b)  &=& U(\bar{b}) b^{3/2}              \label{eq:solution-u1} \\
V_{1,2}(b) &=& V_{1,2}(\bar{b})b^{1/2}    \label{eq:solution-v1} \\ 
\delta(b) &=& b^2 \left[ \delta(\bar{b}) + 4 B_1 u (n+2) T^{1/4} 
                          \right]      \label{eq:solution-delta1} \\ 
D(b) &=& b \left[ D(\bar{b}) + 2 B_1 (nv_1 + v_2) T^{3/4} \right].
                                       \label{eq:solution-d1} 
\eea
Now one can ensure the consistency of using the linearized equations 
by demanding that, when scaling stops at $\delta(b^{\ast})=1$, the 
coupling constant $U$ is small: 
\be 
U(b^{\ast}) \ll 1.                     \label{eq:ginzburg}
\ee 
This condition 
corresponds to the Ginzburg criterion \cite{cardy1} and is violated only 
very close to the line $\delta(b^{\ast})=0$, which defines 
the N\'{e}el temperature $T_N$ as a function of $u$ and $r_1$:
\be
r_1 + (A+4B_1) (n+2) u T^{5/4}_c = 0.  \label{eq:transition-line}
\ee
The Ginzburg criterion (\ref{eq:ginzburg}) is violated 
in a narrow window $\delta T_N/T_N \sim u^{1/3} T^{1/12}_c \ll 1$ 
of strong classical fluctuations. 

To establish connection with experiment, it is assumed that $r_1$ 
and $r_2$ are both proportional to the deviation $(p - p_c)$ 
of the control parameter $p$ from its critical value $p_c$.  
This is a reasonable assumption 
given that the theory at hand is above the upper critical dimension. 
Thus $T_N$ scales as per 
\be
T_N \sim (p_c - p)^{4/5}.  \label{eq:transition-temperature}
\ee
However, which two phases does the line (\ref{eq:transition-temperature}) 
separate? 

\section{The phase diagram}

\begin{figure}
\epsfxsize=6.0 truein
\epsfbox{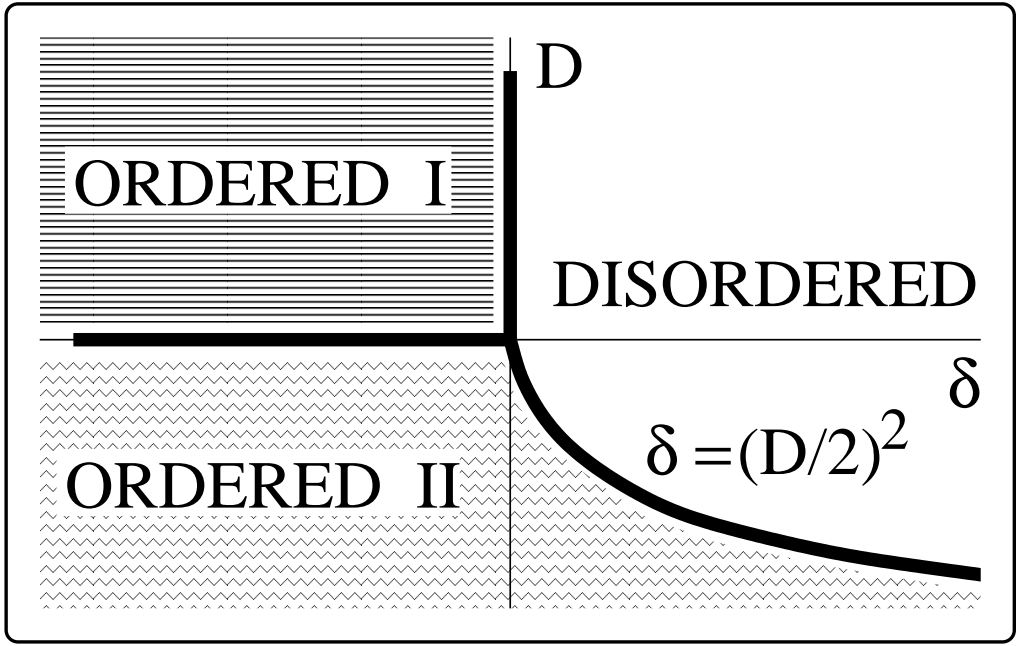}
\caption[Phase diagram near a finite-temperature Lifshitz point.]
{ Schematic phase diagram in the vicinity of 
a Lifshitz point. The ordered phase I corresponds to ordering 
at the wave vector $q=0$ and is shaded by horizontal lines, 
whereas the ordered phase II corresponds 
to ordering at $q= \pm \sqrt{-D/2}$ at $D < 0$ and is marked 
by zig-zag shading. 
}
\label{fig:toy-diagram}
\end{figure}

\begin{figure}
\epsfxsize=6.0 truein
\epsfbox{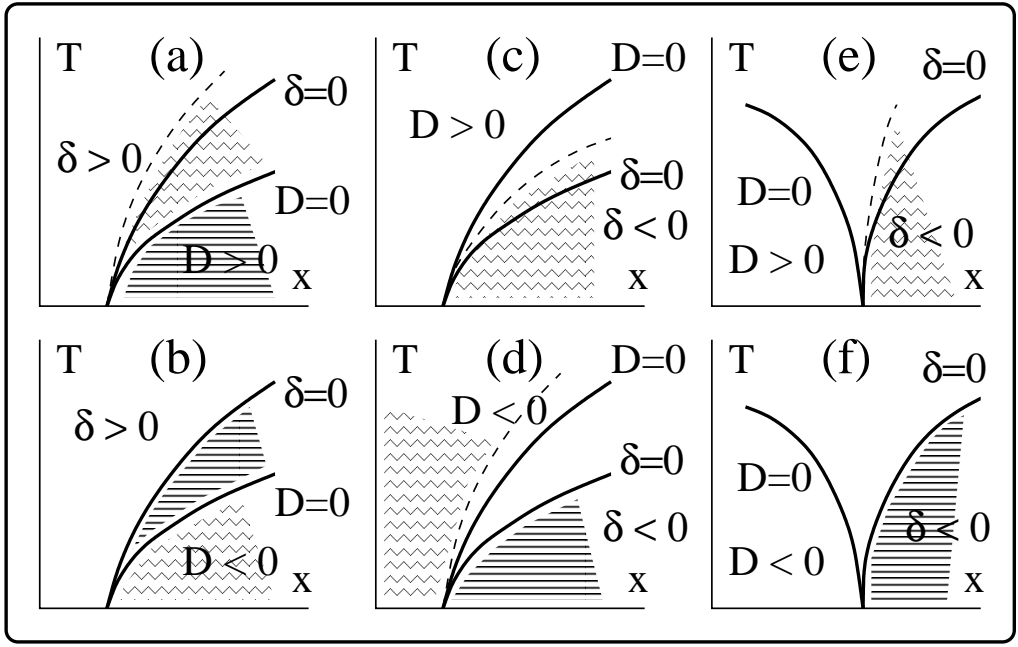}
\caption[Phase diagram near a quantum Lifshitz point.]
{ Schematic phase diagram in the vicinity of 
a quantum Lifshitz point. As in Fig. \ref{fig:toy-diagram}, 
horizontal shading denotes ``commensurate'' ordering 
(at $q=0$), whereas zig-zag shading corresponds to 
``incommensurate'' order (at $q= \pm \sqrt{-D/2}$). 
Wherever shown, dashed line corresponds to $\delta = (D/2)^2$ 
at $D < 0$. The bold lines defined by $D=0$ and $\delta=0$ 
are determined by equations (\ref{eq:solution-delta1}) 
and (\ref{eq:solution-d1}).
}
\label{fig:all-diagram}
\end{figure}

To answer this question, one has to recall that,  
as mentioned above, Lifshitz point is a tricritical point. 
Thus, generally, one shall expect to see two different ordered 
phases and one disordered phase in its vicinity, as illustrated 
by the following toy-model expression for the free energy $F$ 
at a finite-temperature Lifshitz point: 
\be
F \sim \phi \left[ \delta + Dq^2 + q^4 \right] \phi. \label{eq:illustration}
\ee
As follows from (\ref{eq:illustration}), the region $(D > 0, \delta < 0)$ 
corresponds to the phase ordered at $q=0$, whereas in the region 
$(D < 0, \delta < (D/2)^2)$ one finds ordering at wave vectors 
$q = \pm \sqrt{-D/2}$. This phase diagram is shown in Fig. 
\ref{fig:toy-diagram}. 

To establish the phase diagram near a quantum Lifshitz point, one shall 
draw the curves of $\delta(b^{\ast})=0$ and $D(b^{\ast})=0$ in the 
$(T-p)$ plane and compare with Fig. \ref{fig:toy-diagram}. 
Mutual position of the two curves depends on the relative magnitude 
of $u$, $v_1$ and $v_2$, and on the signs of the proportionality 
coefficients between $r_{1, 2}$ and $(p - p_c)$. Six possibilities arise, 
as shown on Fig. \ref{fig:all-diagram}. 
As in Fig. \ref{fig:toy-diagram}, 
horizontal shading denotes ``commensurate'' ordering 
(at $q=0$), whereas zig-zag shading corresponds to 
``incommensurate'' order (at $q= \pm \sqrt{-D/2}$). 
Case (d) can be ruled out on physical grounds, as it corresponds 
to ordering at high temperatures. 
The possibilities illustrated in Fig. \ref{fig:all-diagram} (a) and (b) 
are irrelevant to the results obtained for CePd$_2$Si$_2$ or for 
CeCu$_{6-x}$Au$_x$, as they lead to existence of two ordered states, 
whereas no trace of second transition has been found in either of the 
materials of interest. The remaining cases (c), (e) and (f) 
are the most interesting for us, as they show how a quantum Lifshitz 
point can mimic a ``regular'' quantum critical point with only one 
ordered phase. Case (f)  corresponds to ``commensurate'' ordering, 
while (c) and (e) describe ``incommensurate'' order. 
In the latter two cases, the transition line between the 
``incommensurate'' and the disordered phases is described by 
the equation $\delta (b) = (D(b)/2)^2$, shown in 
Fig. \ref{fig:all-diagram} by dashed line, with $\delta (b)$ and $D(b)$ 
given by (\ref{eq:solution-delta1}-\ref{eq:solution-d1}). At low 
temperatures, this line asymptotically coincides with 
the line $\delta (b)=0$ given by (\ref{eq:transition-temperature}). 

Currently, the structure of magnetic order 
in CeCu$_{6-x}$Au$_x$ is being mapped out experimentally by 
several groups \cite{alert1}. 
However, if a Lifshitz point described above 
is realized in CeCu$_{6-x}$Au$_x$, the exact character of 
ordering is irrelevant for the physical properties 
in the ``classical Gaussian'' region, 
roughly corresponding to the vicinity 
of the line $(T > 0, T_c = 0)$. 

\section{Generation of stiffness} 

\begin{figure}
\epsfxsize=6.0 truein
\epsfbox{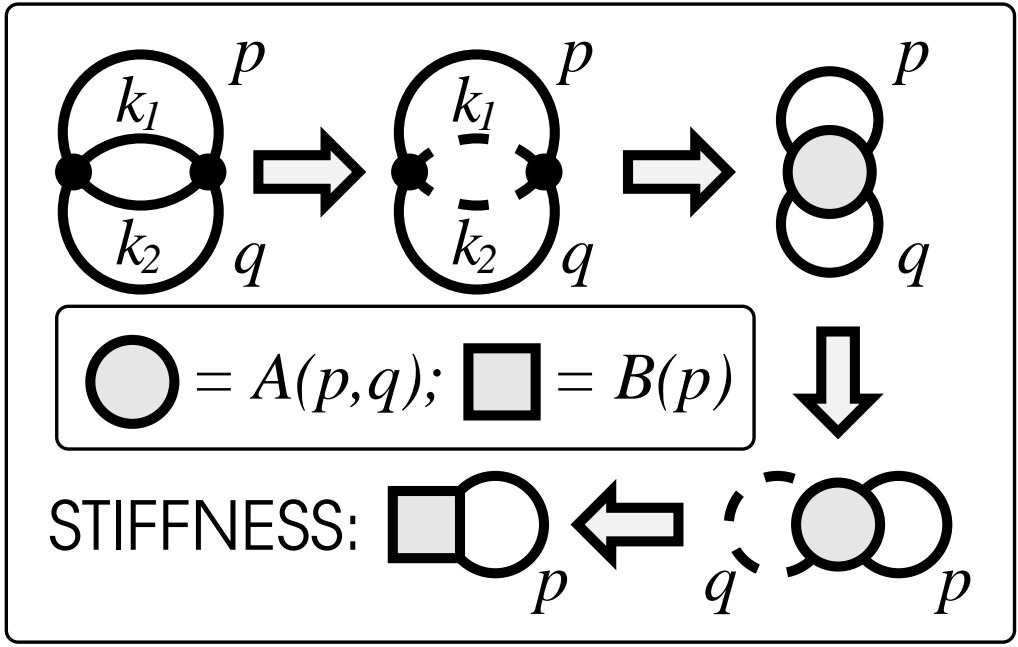}
\caption[Generation of stiffness at a quantum Lifshitz point]
{ Generation of stiffness in the ``soft'' direction
by the renormalization process. Propagators of ``fast'' 
degrees of freedom, integrated out at each step 
of renormalization, are shown by dashed lines. 
The free energy correction in the second order in point-like 
quartic interaction $u$ (black dot) generates 
dispersion of the effective quartic interaction (shaded circle). 
The latter, in turn, gives rise to the stiffness 
in the ``soft'' direction
(shaded square).
}
\label{fig:stiffness}
\end{figure}

As shown in Section III (see Fig. \ref{fig:stiffness}), 
near the quantum Lifshitz point 
the short-range fluctuations do generate stiffness 
in the ``soft'' direction, even though it is absent 
exactly at the critical point. 
To check the importance of this effect at finite temperature, 
one has to compare the generated quadratic term $D q_{\perp}^2$ 
with the quartic term $q_{\perp}^4$. The comparison can be done easily
using the solution (\ref{eq:solution-t}-\ref{eq:solution-d})
of the scaling equations (\ref{eq:t}-\ref{eq:d}).  

At each step of the renormalization procedure, 
one focuses on momenta and frequencies near the current 
value of the cut-off. Recalling the agreement to set the 
cut-off equal to unity, one finds that rescaling factor $b$
corresponds to frequency and momenta $q_{\parallel}(b)=1/b$, 
$q_{\perp}(b)=1/\sqrt{b}$ and $\omega(b) = 1/b^2$. 

Thus the relative importance of the generated term $D(b) q_{\perp}^2(b)$ 
is given by its comparison with the mass term $\delta(b)$ and with 
the quartic term $q_{\perp}^4(b)$. 
To do the comparison in the classical Gaussian region 
which we are studying,
one shall set $r_1 = r_2 = 0$ and then, 
using (\ref{eq:solution-t}-\ref{eq:solution-d}),  
estimate $\delta(b)$ and $D(b)$ at a running value of $b$. 

In the regime of ``quantum'' renormalization ($T(b) \ll 1$), 
the sought estimate is given by 
$$\delta(b) \sim A (n+2) u b^{-1/2}, $$
$$D(b) \sim A (nv_1 + v_2) b^{-3/2}, $$
$$D(b)q_{\perp}^2 (b) \sim A (nv_1 + v_2) b^{-5/2}. $$ 
$$q_{\perp}^4 (b) \sim q_{\parallel}^2 (b) 
\sim |\omega|(b) \sim b^{-2}. $$ 
Applicability of our scaling equations requires weak coupling 
($u, v_{1,2} \ll 1$). Thus, $D(b)q_{\perp}^2 (b)$ is negligible 
compared with $q_{\perp}^4 (b)$ at any value of $b$ 
up to the point when $\delta(b)$ and $q_{\perp}^4 (b)$ become 
of the same order of magnitude. Therefore, the stiffness generated 
by short-distance fluctuations under the renormalization flow 
can be safely neglected in our study, which in turn means that the 
quantum Lifshitz point is well-defined.  

\section{Conclusions}

In this paper, I studied the classical Gaussian region 
near a quantum Lifshitz point. This corresponds to defining the 
low-temperature behavior at the point where the N\'{e}el temperature 
and the stiffness in the ``soft'' direction simultaneously become 
equal to zero. The N\'{e}el temperature was found to scale 
as $T_N \sim (p_c - p)^{4/5}$. The specific heat coefficient was 
found to have a $T^{1/4}$ anomaly, whereas the resistivity was 
shown to exhibit $T^{5/4}$ scaling. 
Of these results, only the resistivity exponent finds experimental 
support (in CePd$_{2}$Si$_{2}$ and CeNi$_{2}$Ge$_{2}$), 
while the predictions for the shape of the transition line 
appears to fail in fitting the experimental data. The situation 
with the specific heat data is not yet entirely certain. 

Regardless of possible reasons, discussed briefly 
in Section II, the failure of simple phenomenology based on 
the assumption of a quantum Lifshitz point appears to be unambiguous 
for CePd$_{2}$Si$_{2}$ and  CeCu$_{6-x}$Au$_{x}$. More generally, 
it appears that the correct theory of the transition shall fall 
below the upper critical dimension, while the Hertz-Millis theory 
and its present refinement are all essentially Gaussian.   

At the moment, the source of the puzzling behavior of these 
Ce-based materials near a zero-temperature transition remains unclear. 
A comprehensive experimental study (specific heat, 
neutron scattering, thermal transport and NMR/NQR/$\mu$SR) might 
help to resolve some of the pressing issues. 
An intriguing possible direction of theoretical research 
would be to study  the yet poorly understood interplay 
between the incipient correlations in the conduction sea 
and the development of the Kondo effect \cite{larnikov1}. 

I am indebted to G. Aeppli, P. Coleman, A. Millis, 
D. Morr, A. Rosch and A. Schr\"{o}der for discussions related 
to this article, and to Y. Aoki and T. Fukuhara for discussions 
of their data. 
The work was started at Rutgers University, 
where it was supported by the National Science 
Foundation (grant number DMR-96-14999). 
Work at the University of Illinois was supported in part 
by the MacArthur Chair endowed by the John D. 
and Catherine T. MacArthur Foundation at the University of Illinois. 

\section*{Appendix}

In this Appendix, I derive the scaling equations (\ref{eq:t}-\ref{eq:d}). 
The first term on the right hand side of each equation 
corresponds to rescaling of the variable under an infinitesimal 
time- and length-scale transformation. It can be obtained e.g. 
by following Hertz \cite{hertz1}.   
Rewrite the action (\ref{eq:action1}-\ref{eq:action3}) 
in the momentum and frequency domain,
replacing the sum over the Matsubara frequencies by an integral 
up to the cut-off (set equal to 1):

\bea
S_{eff}[ \phi ] &=& S^{(2)} + S^{(4)}            \label{eq:action1a}\\
S^{(2)}         &=& 
\int^{1} \frac{d\omega}{2\pi} \int^1 \frac{dq_{\parallel}^2}{(2\pi)^2} 
\int^1 \frac{dq_{\perp}}{2\pi} 
\phi_{\alpha}
\left[
\delta + 
|\omega| + q^{2}_{\parallel}
 + D q^{2}_{\perp} + 
q^{4}_{\perp}
\right]
\phi_{\alpha}                              \label{eq:action2a}\\                                        
S^{(4)}        &=& 
\left[
\int^{1} \frac{d\omega}{2\pi} \int^1 \frac{dq_{\parallel}^2}{(2\pi)^2} 
\int^1 \frac{dq_{\perp}}{2\pi} 
\right]^3_{1,2,3}                           \nonumber \\
 & & \left[
u (\phi_{\alpha}\phi_{\alpha})^2
+ v_1 (q_{\perp}\phi_{\alpha})^2 \phi_\beta \phi_\beta 
+ v_2 
(\phi_{\alpha} q_{\perp} \phi_{\alpha})^2
\right].                                    \label{eq:action3a}   
\eea

Then integrate out a thin shell between the original 
cut-off (equal to 1) and the new cut-off (equal to $1/b$) 
in the $q_{\parallel}$ space. Now define thin shells 
in the $\omega$ and $q_{\perp}$ spaces in such a way that 
upon being integrated out, they would admit rescaling of 
all the variables ($\omega, q_{\parallel}, q_{\perp}, \delta, 
D, u, v_1, v_2$ and $\phi_{\alpha}$) in such a way as to bring 
the remaining part of $S^{(2)}$ back exactly to the form (18) 
with the new values of $\delta, D, u, v_1$ and $v_2$.  
The only choice of rescaling factors which allows this corresponds 
to rescaling $q_{\parallel}$ by $1/b$, integrating out $\omega$ 
between $1$ and $1/b^2$ and rescaling it by $1/b^2$, integrating out 
$q_{\perp}$ between $1$ and $1/\sqrt{b}$ and rescaling it by $1/\sqrt{b}$
 -- at the expense of rescaling $\delta$ and $T$ by $b^2$, $D$ by $b$, 
$u$ by $1/\sqrt{b}$ and $v_{1, 2}$ by $b^{-3/2}$, 
which corresponds precisely to the first 
terms on the right hand side of (\ref{eq:t}-\ref{eq:d}). 

Now, I will obtain the remaining terms in (\ref{eq:delta}-\ref{eq:d}). 
First, expand the 
partition function $Z$ to first order in $S^{(4)}$:  
$$ Z = Z_{0} 
\left[ 1 - \langle S^{(4)} \rangle \right].       
$$
Using Wick's theorem \cite{cardy1}, 
the average $\langle S^{(4)} \rangle$ 
can be conveniently rewritten as 
$$
\langle S^{(4)} \rangle = 
n (n+2) u
\langle \phi^2 \rangle^2
 + n (n v_1 + v_2) \langle \phi^2 \rangle 
\langle q_{\perp}^2 \phi^2 \rangle,         
$$
where $\langle \phi^2 \rangle$ 
and $\langle q_{\perp}^2 \phi^2 \rangle$ are defined as per
\be 
\langle \phi^2 \rangle \equiv 
\int^{1}_0 \frac{dz}{\pi} \coth \left( \frac{z}{2T(b)} \right) 
\int^1 \frac{dq_{\parallel}^2}{(2\pi)^2} 
\int^1 \frac{dq_{\perp}}{2\pi} 
\frac{z}{z^2 + [\delta(b) + q_{\parallel}^2 
+ D(b)q_{\perp}^2 + q_{\perp}^4 ]^2}          \label{eq:average-phi^2}
\ee
\be 
\langle q_{\perp}^2 \phi^2 \rangle \equiv 
\int^{1}_0 \frac{dz}{\pi} \coth \left( \frac{z}{2T(b)} \right) 
\int^1 \frac{dq_{\parallel}^2}{(2\pi)^2} 
\int^1 \frac{dq_{\perp}}{2\pi} 
\frac{z q_{\perp}^2}{z^2 + [\delta(b) + q_{\parallel}^2 
+ D(b)q_{\perp}^2 + q_{\perp}^4 ]^2}.            \label{eq:average-q^2phi^2}
\ee
The next step amounts to integrating out a thin shell near the 
cut-off in the expression for $\langle S^{(4)} \rangle$.  
As described above, such a shell has width $1 - 1/b$ in the 
$q_{\parallel}$ direction, $1 - 1/b^2$ in the $z$ direction 
and $1 - 1/\sqrt{b}$ in the $q_{\perp}$ direction. This integration 
generates the sought terms in (\ref{eq:delta}-\ref{eq:d}), with 
\bea
f_1[T(b), \delta(b), D(b)] &=& 
\frac{1}{2\pi} 
\int^{1}_0 \frac{dz}{\pi} \coth \left( \frac{z}{2T(b)} \right) 
\int^1 \frac{dq_{\perp}}{2\pi} 
\frac{z}{z^2 + [\delta(b) + 1 
+ D(b)q_{\perp}^2 + q_{\perp}^4 ]^2} \nonumber\\
 &+& \frac{2}{\pi} 
\coth \left( \frac{1}{2T(b)} \right) 
\int^1 \frac{dq_{\parallel}^2}{(2\pi)^2} 
\int^1 \frac{dq_{\perp}}{2\pi} 
\frac{1}{1 + [\delta(b) + q_{\parallel}^2 
+ D(b)q_{\perp}^2 + q_{\perp}^4 ]^2} \nonumber \\ 
 &+& \frac{1}{4\pi} 
\int^{1}_0 \frac{dz}{\pi} \coth \left( \frac{z}{2T(b)} \right) 
\int^1 \frac{dq_{\parallel}^2}{(2\pi)^2} 
\frac{z}{z^2 + [\delta(b) +  
q_{\parallel}^2 
+ D(b) + 1 ]^2},                     \nonumber \\
 & &                           \label{eq:f1}
\eea
\bea
f_2[T(b), \delta(b), D(b)] &=& 
\frac{1}{2\pi} 
\int^{1}_0 \frac{dz}{\pi} \coth \left( \frac{z}{2T(b)} \right) 
\int^1 \frac{dq_{\perp}}{2\pi} 
\frac{z}{z^2 + [\delta(b) + 1 
+ D(b)q_{\perp}^2 + q_{\perp}^4 ]^2} \nonumber \\
 &+& \frac{2}{\pi} 
\coth \left( \frac{1}{2T(b)} \right) 
\int^1 \frac{dq_{\parallel}^2}{(2\pi)^2} 
\int^1 \frac{dq_{\perp}}{2\pi} 
\frac{q_{\parallel}^2}{1 + [\delta(b) + q_{\parallel}^2 
+ D(b)q_{\perp}^2 + q_{\perp}^4 ]^2}  \nonumber \\ 
 &+& \frac{1}{4\pi} 
\int^{1}_0 \frac{dz}{\pi} \coth \left( \frac{z}{2T(b)} \right) 
\int^1 \frac{dq_{\parallel}^2}{(2\pi)^2} 
\frac{z q_{\parallel}^2}{z^2 + [\delta(b) +  
q_{\parallel}^2 
+ D(b) + 1 ]^2},                 \nonumber \\
 & &                              \label{eq:f2}
\eea
thus completing the derivation of the scaling equations 
(\ref{eq:delta}-\ref{eq:d}).

\end{document}